\documentstyle[12pt]{article}
\newcommand{\beq}{\begin{equation}}
\newcommand{\eeq}{\end{equation}}
\newcommand{\beqa}{\begin{eqnarray}}
\newcommand{\eeqa}{\end{eqnarray}}
\begin{document}

\baselineskip=24pt

%%%%%%%%%%%%%%%%%%%%%%%%%%%%%%%%%%%%%%%%%%%%%%%%%%%%%%%%%%%%%%%%%%%%%%%%%%%%%
%%%%%%%%%%%%%%%%%%%%%%%%%%%%%% TITLE PAGE %%%%%%%%%%%%%%%%%%%%%%%%%%%%%%%%%%%
%%%%%%%%%%%%%%%%%%%%%%%%%%%%%%%%%%%%%%%%%%%%%%%%%%%%%%%%%%%%%%%%%%%%%%%%%%%%%

\begin{center}
{\bf Generalized Calogero model in arbitrary dimensions}

S.Meljanac$^{a}${\footnote{e-mail: meljanac@irb.hr}}, 
M.Milekovi\'{c} $^{b}$ {\footnote{e-mail: marijan@phy.hr }},
A. Samsarov$^{a}$ {\footnote{e-mail: a.samsarov@irb.hr }}\\ [3mm]

$^{a}$ Rudjer Bo\v{s}kovi\'c Institute, Bijeni\v cka  c.54, HR-10002 Zagreb,
Croatia\\[3mm]

$^{b}$ Theoretical Physics Department, Faculty of Science, P.O.B. 331,
 Bijeni\v{c}ka c.32,
\\ HR-10002 Zagreb, Croatia \\

\end{center}

\setcounter{page}{1}

%%%%%%%%%%%%%%%%%%%%%%%%%%%%%%%%%%%%%%%%%%%%%%%%%%%%%%%%%%%%%%%%%%%%%%%%%%%%%%%

%%%%%%%%%%%%%%%%%%%%%%%%%%%%%%%% ABSTRACT %%%%%%%%%%%%%%%%%%%%%%%%%%%%%%%%%%%%

%%%%%%%%%%%%%%%%%%%%%%%%%%%%%%%%%%%%%%%%%%%%%%%%%%%%%%%%%%%%%%%%%%%%%%%%%%%%%%%%
\begin{center}
\bf Abstract
\end{center}

We define a new   multispecies model of Calogero type in 
D dimensions with harmonic, two-body and three-body interactions. 
Using  the underlying conformal $SU(1,1)$ algebra, 
we indicate how to find the complete set of the states in Bargmann-Fock space.
There are towers of states, with equidistant  energy spectra in each tower. 
We explicitely  construct all polynomial eigenstates, namely the center-of-mass states and 
global dilatation modes, and find their corresponding eigenenergies.
We also construct ladder operators  for these global collective states.
Analysing corresponding Fock space, we detect the universal critical point at which the model 
exhibits singular behavior. The above results are universal for all systems with underlying conformal
$SU(1,1)$ symmetry.

PACS number(s): 03.65.Fd, 03.65.Sq, 05.30.Pr


%%%%%%%%%%%%%%%%%%%%%%%%%%%%%%%%%%%%%%%%%%%%%%%%%%%%%%%%%%%%%%%%%%%%%%%%%%%%%%%%%%%%%%%%%
%%%%%%%%%%%%%%%%%%%%%%%%%%%%%%%%%%%%%%%%%%%%%%%%%%%%%%%%%%%%%%%%%%%%%%%%%%%%%%%%%%%%%%%%%

%%%%%%%%%%%%%%%%%%%%%%%%%%%%%%%%%%%%%%%%%%%%%%%%%%%%%%%%%%%%%%%%%%%

%%%%%%%%%%%%%%%% SECTION 1 : Introduction %%%%%%%%%%%%%%%%%%%%%%%%%

%%%%%%%%%%%%%%%%%%%%%%%%%%%%%%%%%%%%%%%%%%%%%%%%%%%%%%%%%%%%%%%%%%%

\section {Introduction}
The (rational) Calogero model  describes $N$ identical particles  on 
the line which interact through an inverse-square two-body interaction and 
are subjected to a common confining harmonic force. Starting from the inception [1],
the model and its various descendants (also known as Calogero-Sutherland-Moser systems [2]) 
continue to be of interest for both physics and mathematics community, primarly because they are
connected with a number of mathematical and physical problems, ranging from random matrices
and symmetric polynomials [3] to condensed matter systems  and  black hole physics [4]. 
The model is also connected to Haldane's exclusion statistics [5]. The role of Haldane statistical parameter is 
played by (universal) coupling constant in the two-body interaction.
In Haldane's formulation there is the possibility of having particles of different species with 
a mutual statistical coupling parameter depending on the species coupled. This suggest the possible
generalization of the ordinary one-dimensional Calogero model with identical particles to the one-dimensional 
Calogero model with non-identical particles. Distinguishabillity can be introduced by allowing particles to 
have different masses and different couplings to each other. In this way  a one-dimensional multispecies 
Calogero model is obtained [6,7]. \\
Further generalization can be achieved by  formulating the model in dimensions higher than one. 
In the  case of single-species model(s), it was shown  that some exact eigenstates (including the ground state) 
can be obtained for a $D$ dimensions provided that a long-range three-body interaction is added [8]. 
The inevitable appearance of three-body interaction in $D>1$ makes any analysis of such a model(s) 
highly nontrivial and very little is known about their exact solvability.  
Some progress has been achieved only recently for a class of two-dimensional models with identical particles [9].\\
In a present Letter we propose a new type of partially solvable multispecies model of Calogero type in 
D dimensions. In addition to the harmonic potential, it contains  two-body and three-body interactions 
with coupling constants depending on the particle's species. We also allow particles to have different masses.
In this way we incorporate both generalizations mentioned above into a single model. We indicate how to obtain 
(in principle) all eigenstates of the model Hamiltonian in Bargmann representation. 
The  spectrum of states shows a remarkable simplicity. There are towers of states with equidistant energies.
We are able to find all  polynomial eigenstates and corresponding eigenenergies of the Hamiltonian,  
describing global collective states.  
Closer inspection of the Fock space, corresponding to the relative motion of particles, reveals the existence 
of the universal critical point at which system exhibits singular behaviour. This result generalizes that
mentioned in [7,10]. Our results are universal and applicable to all systems with underlying $SU(1,1)$ algebra.

%%%%%%%%%%%%%%%%%%%%%%%%%%%%%%%%%%%%%%%%%%%%%%%%%%%%%%%%%%%%%%%%%%%%%%%%%%%%%%%%%%%%%%%%%%%%%%

%%%%%%%%%%%%%%%%%%%%%%%%%%%%%%%%%%%%%%%%%%%%%% Section 2: A model Hamiltonian...%%%%%%%%%%%%%%%

%%%%%%%%%%%%%%%%%%%%%%%%%%%%%%%%%%%%%%%%%%%%%%%%%%%%%%%%%%%%%%%%%%%%%%%%%%%%%%%%%%%%%%%%%%%%%%

\section{A model Hamiltonian}

We start the analysis with observation that the exact wave functions of the Calogero model are highly correlated. These 
correlations are encoded in the wave functions in the form of a Jastrow prefactor $(x_i-x_j)^{\nu}$ for any pair
of particles $i,j$. The exponent of the correlator is related to the strength of the two-body interaction. 
It is then plausible to make an ansatz for the most general ground state wave function for the $N$ 
$distinguishable$ Calogero-like particles in $D$ dimensions in the form ($\hbar =1$)  

\beq
 \Psi_{0}(\vec{r}_{1},...,\vec{r}_{N})=\Delta e^{-\frac{\omega}{2}
 \sum_{i=1}^{N} m_{i} {\vec{r}}_{i}^{2}}
\eeq

where the Jastrow prefactor is generalized to

\beq
\Delta= \prod_{i<j}{|\vec{r}_{i}-\vec{r}_{j}|}^{\nu_{ij}},\;\;\;\;
 \nu_{ij}=\nu_{ji},\quad i,j =1,2,\cdots N.
\eeq
Here, $m_{i}$ are masses of the particles, $\omega$ is the frequency of the harmonic 
potential and  $\nu_{ij}$ are the statistical parameters between particles $i$ and $j$. 
In principle, one could start with any wave function with no nodes, except at the coincidence points, and which is
continuosly connected with Gausss function when parameters $\nu_{ij}\rightarrow 0$. 
Note that for  $\nu_{ij}=\nu $, $m_i=m$ and $D\neq 1$, Eq.(1) smoothly goes to exact ground state of the 
Calogero-Marchioro model [11], so the wave function (1) is a  natural choice.
 Adopting the
reasoning from Ref.[12], we can ask for what kind of Hamiltonian is the wave function (1) 
the exact ground state. 
It turns out that  $\Psi_{0}(\vec{r}_{1},...,\vec{r}_{N})$ will be, for sufficiently small deformations $\nu_{ij}$,
the exact ground state of the Hamiltonian
$$
H=-\frac{1}{2}\sum_{i=1}^{N}\frac{1}{m_{i}} \vec{\nabla}_{i}^{2}+ 
\frac{{\omega}^{2}}{2} \sum_{i=1}^{N} m_{i} \vec{r}_{i}^{2}
$$
$$
 + \frac{1}{2} \sum_{i<j} \frac{\nu_{ij}(\nu_{ij}+D-2)}{{|\vec{r}_{i}-\vec{r}_{j}|}^{2}} (\frac{1}{m_{i}} + 
 \frac{1}{m_{j}})
$$
\beq
+ \frac{1}{2} \sum_{i \neq j, i\neq k} \frac{\nu_{ij} \nu_{ik} (\vec{r}_{i}-\vec{r}_{j}) (\vec{r}_{i}-\vec{r}_{k})}{m_{i}  	
{|\vec{r}_{i}-\vec{r}_{j}|}^{2} {|\vec{r}_{i}-\vec{r}_{k}|}^{2}},
\eeq
such that 
\beq
H \Psi_{0} = E_{0} \Psi_{0}, 
\eeq
\beq
E_{0} = \omega ( \frac{N D}{2} + \sum_{i<j} \nu_{ij})\equiv \omega  \epsilon_0 .
\eeq
The ground state (1) and the Hamiltonian (3) are invariant under the group of permutation of $N$ elements, $S_N$, 
generated by exchange operators $K_{ij}$ [13]. Operators $K_{ij}$ interchange indices $i \leftrightarrow j $ in 
all quantities, i.e.
$m_{i} \leftrightarrow m_{j} ,\; \nu_{ik} \leftrightarrow \nu_{jk} ,
\; \vec{r}_{i} \leftrightarrow \vec{r}_{j} ,\; \vec{p}_{i} \leftrightarrow \vec{p}_{j}$.\\
For $D=1$ the three-body term in (3) identically vanish if  $\nu_{ij}=\nu$,
  $m_i=m$ or if
$\nu_{ij}=\alpha m_i m_j $,  $\alpha $ beeing some universal constant [7]. 
Unlike in one dimension, however,  it does not
vanish in higher dimensions and plays a crucial role in the analysis that is to follow.

Let us perform the non-unitary  transformation on $\Psi_{0}$, namely 
$\tilde{\Psi}_{0} = \Delta^{-1} \Psi_{0}$. It generates a similarity transformation which leads to another
$S_N$ invariant Hamiltonian $\tilde{H} = \Delta^{-1}H \Delta $. We find $\tilde{H}$ as

$$
\tilde{H} = -\frac{1}{2} \sum_{i=1}^{N} \frac{1}{m_{i}} {\vec{\nabla}_{i}}^{2}+ 
\frac{{\omega}^{2}}{2} \sum_{i=1}^{N} m_{i} {\vec{r}}_{i}^{2}  
-\sum_{i<j} \nu_{ij}\frac{(\vec{r}_{i}-\vec{r}_{j})}  {{|\vec{r}_{i}-\vec{r}_{j}|}^{2}} 
(\frac{1}{m_{i}} \vec{\nabla}_{i} - \frac{1}{m_{j}} \vec{\nabla}_{j}) 
$$
\beq
 =   {\omega}^{2} T_{+} - T_{-}.
\eeq
where we have introduced
\beq
T_{-}=  \frac{1}{2} \sum_{i=1}^{N} \frac{1}{m_{i}} {\vec{\nabla}_{i}}^{2} +  
\sum_{i<j} \nu_{ij}\frac{(\vec{r}_{i}-\vec{r}_{j})}  {{|\vec{r}_{i}-\vec{r}_{j}|}^{2}} 
(\frac{1}{m_{i}} \vec{\nabla}_{i} - \frac{1}{m_{j}} \vec{\nabla}_{j}),
\eeq
$$
T_{+}= \frac{1}{2} \sum_{i=1}^{N} m_{i} {\vec{r}}_{i}^{2}, \qquad \qquad 
T_{0} = \frac{1}{2} (\sum_{i=1}^{N} \vec{r}_{i} \vec{\nabla}_{i} + 
{\varepsilon}_{0}).                   
$$
The operators $T_{\pm} ,\; T_{0}$ satisfy the $ SU(1,1) $ algebra
\beq
[T_{-} , T_{+}] = 2T_{0}, \qquad [T_{0},T_{\pm}] = \pm T_{\pm}.
\eeq
%%%%%%%%%%%%%%%%%%%%%%%%%%%%%%%%%%%%%%%%%%%%%%%%%%%%%%%%%%%%%%%%%%%%%%%%%%%%%%%%%%%%%%%%%%%%%%%%%%%%%%%%%%%%%
The following identity (i.e. similarity transformation) holds for $\omega \neq 0$:
\beq
\tilde{H} = {\omega}^{2} T_{+} - T_{-} = 2 \omega S \, T_0 \, S^{ -1},
\eeq
$$
S= e^{-\omega T_{+}} \, e^{-\frac{1}{2\omega} T_{-}}.
$$
Owing to this identity, we can employ Bargmann representation and construct iteratively Bargmann-Fock space 
of eigenstates. We begin with state $\Phi_0$, which is  the lowest weight vector of the operator $T_{-}$ and also
an eigenstate of $T_{0}$:
\beq
T_{-} \Phi_0 = 0, \qquad T_{0} \Phi_0 = \frac{\epsilon_0}{2} \Phi_0.
\eeq
In our case, $\Phi_0 = 1$ and $\epsilon_0$ is given in Eq.(5). \\
The tower of excited states ( level 0-tower ) is obtained by succesive application of $T_{-}$ operator:
\beq
T_{-}\Phi_{2p}=\Phi_{2p-2}, \qquad    2\omega T_0 \Phi_{2p}= \omega (2 p + \epsilon_0) \Phi_{2p} ,  \qquad p=0,1,2,3....
\eeq
The states $\Phi_{2p}$ are either polynomials or irrational functions of homogenity $2p$, 
and are eigenstates of $T_{0}$. 
Two succesive  states differ in energy by an amount $2\omega$.\\
Similarly, one can construct towers of  states at level 1, $\Phi^{I_1}_{2p+1}$, $p=0,1,2...,\\
I_1=(i_1=1,2,...N;\alpha_1=1,2,...D)$, using
\beq
T_{-}\Phi^{I_1}_{1}=0, \qquad T_{-}\Phi^{I_1}_{2p+1}= \Phi^{I_1}_{2p-1}, \qquad 
2\omega T_0 \Phi^{I_1}_{2p+1}= \omega (2 p + \epsilon_1^{I_1}) \Phi^{I_1}_{2 p + 1}.
\eeq
Here, $\epsilon_1^{I_1}$ is energy of the first excited  state which tends to $(1+\frac{ND}{2})$ in the limit
$\nu_{ij}\rightarrow 0$. 
Two succesive  states also differ in energy by an amount $2\omega$. \\
Following the procedure, one gets the towers of states at level $k $, $0 \leq k \leq N D$, using
$$
T_{-}\Phi^{I_1,...,I_k}_{k}=0, \qquad T_{-}\Phi^{I_1,...,I_k}_{2p+k}= \Phi^{I_1,...,I_k}_{2p +k -2},\qquad 
2\omega T_0 \Phi^{I_1,...,I_k}_{2p+k}= \omega (2 p + \epsilon_k^{I_1,...,I_k}) \Phi^{I_1,...,I_k}_{2 p + k}.
$$
Here, the energies $\epsilon_k^{I_1,...,I_k}$ tends to $(k +\frac{ND}{2})$ in the limit $\nu_{ij}\rightarrow 0$.
The states $\Phi^{I_1,...,I_k}_{2p+k}$ are eigenstates of the Hamiltonian $2\omega T_{0}=S^{ -1}\, \tilde{H}\, S
$, Eq.(9). Particularly,  the state $\Phi_0 = 1$ is a ground state (i.e. the lowest energy eigenstate) for all 
towers if $\epsilon_0 < \epsilon_k^{I_1,...I_k}$, $\forall I_1,...,I_k$ and for all indices $k$.\\
 Notice that the operator $T_{+}$ of Eq.(7), acting on the particular state in the given tower, 
gives an another  state in the same tower with energy greater by an amount $2\omega$ (see also Sec. 3, Eq.(17)).\\
The procedure outlined above in Eqs.(10-12) is exaustive, i.e. it gives all eigenstates of the  $S$-transformed
Hamiltonian  $S^{ -1}\, \tilde{H}\, S$ (cf. Eq.(9)) , 
provided one is able to solve  differential equations (11) and (12). This is a non-trivial task, even in $D=1$. 
However, one can readily show that this procedure, when applied to the system of $N$ D-dimensional free harmonic
oscillators ($\nu_{ij}=0$ in Eqs.(6) and (3)) yields  the following set of eigenstates for $\tilde{H}=H$:
\beq
S \, \cdot \, ( \prod_{i=1}^N \prod_{\alpha =1}^D  r_{i,\alpha}^{n_{i,\alpha}} )  , 
\qquad n_{i,\alpha}=0,1,2..., ,
\eeq
where (cf. Eqs.(7) and (9))
$$
S=e^{-\omega T_{+}} \, e^{-\frac{1}{2\omega} T_{-}}=  
e^ {-\omega\frac{1}{2} \sum_{i=1}^{N} m_{i} {\vec{r}}_{i}^{2}}\, 
e^{-\frac{1}{2\omega} \frac{1}{2} \sum_{i=1}^{N} \frac{1}{m_{i}} {\vec{\nabla}_{i}}^{2} }.
$$  
The corresponding  eigenenergies are
\beq
\omega ( \frac{ND}{2} + \sum_{i=1}^N \sum_{\alpha =1}^D n_{i,\alpha} ).
\eeq
For the convenience of the reader, we describe in the next Table the structure of the few lowest towers of 
states at level $k$ (Eqs.(10)-(12)) in this simple case.
$$
\begin{tabular}{|c|c|c|} \hline
$k$ & level-$k$ tower & indices\\ \hline \hline
0   & 1 & -\\ \hline \hline
1   & $r_{I}$ & $I={(i,\alpha)}$ \\ \hline \hline
2   & $r_{I_1}r_{I_2}$ & $I_1\neq I_2$ \\ \hline \hline
3   &  $r_{I_1}r_{I_2}r_{I_3}$ & $I_1\neq I_2\neq I_3\neq I_1$\\ \hline \hline
$\vdots$ & $\vdots$ & $\vdots$   \\ \hline %\hline

\end{tabular}
$$
For the general case, Eqs.(10)-(12), the towers of states at level $k$, $\Phi^{I_1,...,I_k}_{k} $, need not
have such a simple monomial structure since they can be , in principle, homogenious irrational functions.
From Eq.(13) one can count the number of states at each level of given homogenity. For example, 
there are $ND$ states of homogenity one, $ND + \frac{ND \,(ND-1)}{2}$ states of homogenity two, etc. There are
$2^{ND}$ towers in total.\\
Now, one can put  an interesting question, namely is there, for sufficiently small deformation parameters $\nu_{ij}$,
one-to-one correspondence
between our multispecies model $\tilde{H}(\nu_{ij})$, Eq.(6), and $N$ D-dimensional free harmonic oscillators \\
$H(\nu_{ij}=0)=\tilde{H}(\nu_{ij}=0)$. According to our analysis, there is no unique similarity transformation between these two
systems. However, there is similarity transformation between given tower in the interacting system 
($\nu_{ij}\neq 0$) and analogous tower in the free system ($\nu_{ij}= 0$), up to the constant 
$ \epsilon_k^{I_1,...I_k}$. Particularly, this was shown  for $D=1$ and identical particles ($\nu_{ij}=\nu $) in
Ref.[9]. In that case, the eigenstates are restricted to $S_N$-symmetric representations.

We are unable to find towers of states by solving differential equations (11-12) in general. 
However, as we will show in the
next section, we are able to construct global collective states for the Hamiltonian (6). These states represent all states of the
polynomial type in Bargmann representation in generic case. Moreover, these states are universal for all systems with
underlying conformal $SU(1,1)$ symmetry.

%%%%%%%%%%%%%%%%%%%%%%%%%%%%%%%%%%%%%%%%%%%%%%%%%%%%%%%%%%%%%%%%%%%%%%%%%%%%%%%%%%%%%
%%%%%%%%%%%%%%%%%%%%%%%%%%%%%%%%  Section 3: Ladder operators.....%%%%%%%%%%%%%%%%%
%%%%%%%%%%%%%%%%%%%%%%%%%%%%%%%%%%%%%%%%%%%%%%%%%%%%%%%%%%%%%%%%%%%%%%%%%%%%%%%%%%%%%+
\section{Ladder operators and Fock space representation for global collective states}

It is convenient to introduce the center-of-mass coordinate $ \vec{R} $ and 
the relative coordinates $ \vec{\rho}_{i} $ [14]:
$$
\vec{R} = \frac{1}{M}  \sum_{i=1}^{N} m_{i} \vec{r}_{i}, \qquad \vec{\nabla}_{R} = \sum_{i=1}^{N}  \vec{\nabla}_{i}
$$
\beq
\vec{\rho}_{i} = \vec{r}_{i} - \vec{R}, \qquad \vec{\nabla}_{\rho_i}=
\vec{\nabla}_{i}-\frac{m_i}{M}\vec{\nabla}_{R}
\eeq
They satisfy identities $\sum_{i=1}^{N} m_{i} \vec{\rho}_{i} =   \sum_{i=1}^{N} 
 \vec{\nabla}_{\rho_{i}} = 0.$
In terms of the variables just introduced, the Hamiltonian $\tilde{H}$ and wave function
$\tilde{\Psi}_{0}$ separate into parts which describe  center-of-mass motion (CM) and  relative motion (R),
namely $\tilde{H}=\tilde{H}_{CM} + \tilde{H}_{R}$ and 
$\tilde{\Psi}_{0}(\vec{r}_{1},...,\vec{r}_{N})=
\tilde{\Psi}_{0}(\vec{R})
\tilde{\Psi}_{0}(\vec{\rho}_{1},...,\vec{\rho}_{N})$.\\
Using Eqs.(7) and (15) we define creation (+) and annihilation (-) operators 
$$
\vec{A_{1}}^{\pm} = \frac{1}{\sqrt{2}} (\sqrt{M \omega} \vec{R} \mp \frac{1}{\sqrt{M \omega}} \vec{\nabla}_{R}),
$$
\beq
{A_{2}}^{\pm} = \frac{1}{2} ( \frac{T_{-}}{\omega} + \omega T_{+}) \mp T_{0},
\eeq
which satisfy the following commutation relations ($\alpha, \beta =1,2,\cdots
D$) :
$$
[A_{1,\alpha}^{-}, A_{1,\beta}^{+}] = \delta_{\alpha\beta},\qquad 
 [A_{1,\alpha}^{-}, A_{1,\beta}^{-}] = 
[A_{1,\alpha}^{+}, A_{1,\beta}^{+}] = 0, 
$$
$$
[\vec{A_{1}}^{-}, {A_{2}}^{+}] = \vec{A_{1}}^{+}, \qquad
[A_{2}^{-},\vec{A_{1}}^{+}] = \vec{A_{1}}^{-},
$$
$$
[A_{2}^{-}, {A_{2}}^{+}] = \frac{\tilde{H}}{\omega}, \qquad 
[\tilde{H}, \vec{A_{1}}^{\pm}] = \pm \omega \vec{A_{1}}^{\pm} ,\qquad 
$$
\beq
[\tilde{H}, {A_{2}}^{\pm}] = \pm 2 \omega{A_{2}}^{\pm}.
\eeq
Notice that ${A_{2}}^{\pm}= S\, T_{\pm}\, S^{-1}$, $\vec{A_{1}}^{+}= S\,\vec{R}\, S^{-1}$ and 
$\vec{A_{1}}^{-}= S\, \vec{\nabla}_{R} \, S^{-1}$, with $S$ defind in Eq.(9).
They act on the Fock vacuum $|\tilde{0} \rangle \propto \tilde{\Psi}_{0}(\vec{r}_{1},...,\vec{r}_{N})$ as
\beq
\vec{A_{1}}^{-}|\tilde{0} \rangle = A_{2}^{-}|\tilde{0} \rangle = 0 , \qquad  \langle\tilde{0} |\tilde{0} \rangle = 1.
\eeq
The excited states in the Fock space, corresponding to global collective states, are of the form  
\beq
{A_{1,1}^{+}}^{n_{1,1}}\cdots {A_{1,D}^{+}}^{n_{1,D}} {A_{2}^{+}}^{n_{2}}|\tilde{0} \rangle \equiv
\prod_{\alpha=1}^{D}(A_{1,\alpha}^{+})^{n_{1,\alpha}} {A_{2}^{+}}^{n_{2}}|\tilde{0} \rangle,
\eeq
where $n_{1,\alpha}=0,1,2... (\forall \alpha)$ and $n_2=0,1,2...$\\
The repeated action of the operators ${A^{+}_{1,\alpha}}$ 
 on the vacuum $|\tilde{0} \rangle $ reproduces, in the coordinate representation,
Hermite polynomials $H_{n_{1,\alpha}}(R_{\alpha}{\sqrt {M\omega}})$. Similarly, the repeated  
action of the operator $ A_{2}^{+}$ on the vacuum $|\tilde{0} \rangle $ reproduces  hypergeometric function, which reduces to 
associated Laguerre polynomials 
$  L_{n_2 + \varepsilon_{0}-1}^{\varepsilon_{0}-1} (2 \omega T_{+}$) for certain values of parameters.
The states (19) are eigenstates of the $\tilde{H}$ with the energy eigenvalues (cf. last two equations in Eqs.(17))
\beq
E_{n_{1,\alpha}; n_{2}} = \omega ( \sum_{\alpha = 1}^{D} n_{1, \alpha} + 2n_{2} + {\varepsilon}_{0}).
\eeq
This is the part of the complete spectrum  which corresponds to center-of-mass states and global dilatation states,
respectively. 

Now we show that the states (19) are perfectly normalizable,i.e. quadratically integrable and physically acceptable for both
Hamiltonians $\tilde{H}$ and $H$, provided that $\epsilon_0 >\frac{D}{2}$.
First, we completely decouple CM- and R-motion by introducing another set of the creation and annihilation operators
$\{B_{2}^{+},B_{2}^{-}\}$:  
\beq
B_{2}^{\pm} = {A_{2}}^{\pm} - \frac{1}{2}(\vec{A_{1}}^{\pm})^{2} ,
\eeq
 such that
\beq
[{A_{1, \alpha}}^{\pm}, {B_{2}}^{\mp}] = 0 .
\eeq
Hence, we get
$$
\tilde{H}_{R}= \omega [B_{2}^{-},B_{2}^{+}] ,\qquad
 [\tilde{H}_{R},B_{2}^{\pm}] = \pm 2 \omega B_{2}^{\pm},
$$
\beq
\tilde{H}_{CM}=\frac{1}{2}\omega \sum_{\alpha = 1}^{D} \{ A_{1, \alpha}^{-} , 
{A_{1, \alpha}}^{+} \}_+ .
\eeq
The Fock space now splits  into the CM-Fock space, spanned by 
$\prod_{\alpha=1}^{D}(A_{1,\alpha}^{+})^{n_{1,\alpha}} |\tilde{0} \rangle_{CM}$ and the R-Fock space, spanned by
$B_{2}^{+ n_2}|\tilde{0}\rangle_{R} $, where 
$|\tilde{0}\rangle_{CM}\propto  e^{-\frac{\omega}{2} M {\vec{R}}^{2}} $ and 
$ |\tilde{0}\rangle_{R} \propto e^{-\frac{\omega}{2} 
\sum_i m_{i} \vec{\rho}_{i}^{2}} $. We point out that the R-modes are universal for all systems with underlying conformal
$SU(1,1)$ symmetry, i.e. for the Hamiltonians of the form $H= -T_{-} + \omega^2 T_{+} +\gamma T_{0}$, where 
$T_{\pm}, T_{0}$ satisfy $SU(1,1)$ algebra (8).\\
Closer inspection of the R-Fock space of the Hamiltonian $\tilde{H}_{R}$, Eq.(23), reveals the existence of the 
universal critical point defined by  the  zero-energy condition
\beq
E_{0R}= \frac{(N-1)D}{2} + \frac{1}{2}\sum_{i\neq j } \nu_{ij}  = 0.
\eeq
At the critical point  the system described by  $\tilde{H}_{R}$  collapses completely. 
This means that the relative coordinates, the relative momenta and the relative energy  are all zero at this 
critical point. There survives only one oscillator, describing the motion of the
centre-of-mass.  Such behaviour resembles some features of the Bose-Einstein condensate. It  was first noticed 
in Ref.[9] for the case $D=1$, $ \nu_{ij}=\nu$ and $m_i=m$. In that case the critical point (24) 
is simply at $\nu=-\frac{1}{N} $. (Notice that there is also critical point at  $ \nu=1+\frac{1}{N}$ for this case).  
Of course, for the initial Hamiltonian $H$, Eq.(3), which is not unitary 
(i.e. physically) equivalent to  $\tilde{H}$, this corresponds to
 some $\nu_{ij} < 0$, satsfying Eq.(24), and the norm of the wave function (1) blows up at the critical point. 
For $\nu_{ij}$ negative but greater than the critical values (24), the wave function is singular 
at coincidence points but still quadratically integrable. Out of the critical point we have one-to-one correspondence
between our multispecies system (6) and the system of $N$ D-dimensional free oscillators, at least for the 
dilatation states $B_{2}^{+ n_2}|\tilde{0}\rangle_{R} $.

%%%%%%%%%%%%%%%%%%%%%%%%%%%%%%%%%%%%%%%%%%%%%%%%%%%%%%%%%%%%%%%%%%%%%%%%%%%%%%%%%%%%%
%%%%%%%%%%%%%%%%%%%%%%%%%%%%%%%%  Section 4: Conclusion.....%%%%%%%%%%%%%%%%%%%%%%%%
%%%%%%%%%%%%%%%%%%%%%%%%%%%%%%%%%%%%%%%%%%%%%%%%%%%%%%%%%%%%%%%%%%%%%%%%%%%%%%%%%%%%%
\section{Conclusion}

In summary, we have defined a nontrivial many-body Hamiltonian $H$ (Eq.(3)) of Calogero type in D dimensions with two- and
three-body interactions among non-identical particles. Strength of the interactions, $\nu_{ij}$, depends on the particle's
species and this feature makes any analysis of such a model  nontrivial, even in $D=1$.   
Using underlying $SU(1,1)$ structure of the transformed Hamiltonian $\tilde {H}$ (Eq.(6)) 
and Bargmann representation we outlined a procedure which gave in principle all eigenstates of the Hamiltonian.
While we were unable to solve corresponding differential equations (11,12), we were able to find some
general features of the solutions.
There are towers of  states with equidistant energy spectra.
 In each tower two neighbouring states differ in energy by $2 \omega$.
Moreover, we managed to  solve $\tilde {H}$ partially, i.e. 
we explicitely found its global collective states,  corresponding to the  center-of-mass motion and 
the relative motion of particles. Those are all polynomial solution in the Bargmann representation in generic case. 
We also found their eigenenergies. The spectrum of collective modes, Eq.(20), is linear, equidistant and degenerate.
It is also found that, for $\sum_{i\neq j } \nu_{ij} = - (N-1)D$, the  
Fock space, corresponding to the relative motion of particles, contained states of zero norm and the whole system 
exhibited singular behaviour. At this critical point the ground state wave function of the Hamiltonian $H$, Eq.(3), posseses infinite norm.\\
If we consider identical Bose (Fermi) particles, with $m_i=m$, and $\nu_{ij}=\nu$, the eigenstates are 
restricted to $S_N$-symmetric (antisymmetric) functions and the critical point is at $\nu=- \frac{D}{N}$. Our 
analysis of multispecies Calogero model gives deeper insight on the single-species Calogero models in higher
dimensions.\\ 
All results presented here are common and universal for all systems with underlying conformal $SU(1,1)$ symmetry.
The potentially most interesting applications of our results might be in two dimensions and quantum Hall efect.

\bigskip
\bigskip
{\bf Acknowledgment}\\
This work was supported by the Ministry of Science and Technology of the Republic of Croatia under 
contracts No. 0098003 and No. 0119261.
\newpage
%%%%%%%%%%%%%%%%%%%%%%%%%%%%%%%%%%%%%%%%%%%%%%%%%%%%%%%%%%%%%%%%%%%%%
%%%%%%%%%%%%%%%%%%%%%%%%%%%%% REFERENCES %%%%%%%%%%%%%%%%%%%%%%%%%%%%
%%%%%%%%%%%%%%%%%%%%%%%%%%%%%%%%%%%%%%%%%%%%%%%%%%%%%%%%%%%%%%%%%%%%%

\end{document}